\documentclass{article}

\usepackage{format/arxiv}
\usepackage{format/richard}
\usepackage{format/richard_category}

\usepackage{authblk}

\title{From Augmentation to Symbiosis: A Review of Human-AI Collaboration Frameworks, Performance, and Perils}
\author[1,2]{Richard Jiarui Tong}

\affil[1]{NEOLAF Inc., NJ, USA}
\affil[2]{IEEE Artificial Intelligent Standards Committee, IEEE, NJ, USA}

\begin{document}
\maketitle
\begin{abstract}
This paper offers a concise, 60-year synthesis of human-AI collaboration, from Licklider's ``man-computer symbiosis" (AI as colleague) and Engelbart's ``augmenting human intellect" (AI as tool) to contemporary poles: Human-Centered AI's ``supertool" and Symbiotic Intelligence's mutual-adaptation model. We formalize the mechanism for effective teaming as a causal chain: Explainable AI (XAI) → co-adaptation → shared mental models (SMMs). A meta-analytic ``performance paradox" is then examined: human-AI teams tend to show negative synergy in judgment/decision tasks (underperforming AI alone) but positive synergy in content creation and problem formulation. We trace failures to the algorithm-in-the-loop dynamic, aversion/bias asymmetries, and cumulative cognitive deskilling. We conclude with a unifying framework—combining extended-self and dual-process theories—arguing that durable gains arise when AI functions as an internalized cognitive component, yielding a unitary human-XAI symbiotic agency. This resolves the paradox and delineates a forward agenda for research and practice.
\end{abstract}

% \tableofcontents

\newpage

% \tableofcontents

% % \clearpage
% \pagenumbering{arabic} % Arabic numerals for main content
% \setcounter{page}{1}

% ------------------------------------------------------------------
% --- Section 1: THE SYMBIOTIC VISION
% ------------------------------------------------------------------
\section{The Symbiotic Vision: Foundational Concepts in Human-Machine Partnership}

\subsection{Introduction: The Licklider Vision of "Man-Computer Symbiosis"}

The conceptual origin of modern human-AI collaboration can be traced to J.C.R. Licklider's 1960 seminal paper, "Man-Computer Symbiosis" \cite{licklider1960}. Decades before the technical capabilities existed, Licklider articulated a vision not of automation, but of deep, co-equal partnership. His framework was a direct response to the limitations of computing at the time, which could only execute pre-formulated problems—a process Licklider found slow and cumbersome, likening it to a "battle... over before the second step in its planning was begun" if one had to wait for a 20-foot-long printout of numbers just to plan the next step.

Licklider's solution was a "cooperative interaction" defined by "very close coupling between the human and the electronic members" of the partnership. This "symbiosis" was envisioned as a "living together in intimate association... of two dissimilar organisms". This biological metaphor was deliberate, setting it apart from mere tool use.

The primary aims of this symbiosis were twofold and remain profoundly relevant today:
\begin{enumerate}
    \item \textbf{To facilitate formulative thinking.} Licklider's primary goal was to "let computers facilitate formulative thinking as they now facilitate the solution of formulated problems". He saw the computer's potential not just as a calculator, but as a partner in the messy, creative, and often "intuitively guided trial-and-error" process of *defining* the problem in the first place.
    \item \textbf{To enable cooperative decision-making.} The second goal was "to enable men and computers to cooperate in making decisions and controlling complex situations" in real-time, "without rigid reliance on predetermined programs".
\end{enumerate}

In this proposed partnership, the division of labor was clear and complementary. Humans would "set goals, form hypotheses, determine criteria, and perform evaluations." Computers, in turn, would handle the "routine work" necessary to "prepare the way for insights and decisions".

This vision's most radical component was its governing metaphor. Licklider was not describing a better tool; he was describing a "colleague." He explicitly stated the need "to think in interaction with a computer in the same way that you think with a colleague whose competence supplements your own". This "teammate" metaphor is the conceptual seed of a 60-year debate that defines the entire field of human-AI interaction. Licklider's hope was that this partnership would "think as no human brain has ever thought". For decades, conventional logic algorithms failed to fulfill this vision, but recent analyses suggest that modern machine learning, particularly deep learning, may represent the first technical architectures capable of achieving the "true Man-Computer Symbiosis" Licklider described.

\subsection{The Engelbart Framework: "Augmenting Human Intellect"}

Independently, and only two years after Licklider's paper, Douglas Engelbart provided a complementary, and arguably competing, foundational vision in his 1962 report, "Augmenting Human Intellect: A Conceptual Framework" \cite{engelbart1962}. While Licklider's vision was philosophical and organismic, Engelbart's was a structured, systems-engineering program for systematically enhancing human capability.

Engelbart's stated goal was "increasing the capability of a man to approach a complex problem situation, to gain comprehension to suit his particular needs, and to derive solutions to problems". Like Licklider, he was focused on solving previously "insoluble" problems and achieving "more-rapid comprehension" and "better comprehension".

However, Engelbart's "Conceptual Framework"  was not a "symbiont"; it was an integrated system. He introduced the "H-LAM/T" system, which conceptualized augmentation as an integrated set of components that must be co-developed:
\begin{itemize}
    \item \textbf{H} (Human): The operator, whose "basic information-handling capabilities" are the subject of augmentation.
    \item \textbf{L} (Language): The way the human structures concepts.
    \item \textbf{A} (Artifacts): The physical tools, such as computers and displays, that "interact with... the individual".
    \item \textbf{M} (Methodology): The procedures and "new methods of thinking and working" that allow the human to capitalize on the system.
    \item \textbf{T} (Training): The process of learning to use the new, integrated system effectively.
\end{itemize}

This framework  moved beyond Licklider's metaphor to an engineering-driven program. The "augmentation means" were the "amplifier" of the human's intelligence. This vision famously led to "The Mother of All Demos" in 1968, showcasing hypertext, real-time text editing, and the computer mouse—all "artifacts" designed to augment the human intellect.

\subsection{Synthesis: The Foundational Tension (Tool vs. Teammate)}

The subtle, yet profound, differences between the visions of Licklider and Engelbart created the central and most enduring tension in the fields of human-computer interaction (HCI) and artificial intelligence: \textbf{Are we building teammates or tools?}

This question, which can be seen as the original schism of the field, dictates all subsequent design philosophies, governance structures, and even research questions.
\begin{itemize}
    \item \textbf{Licklider's Path (The Teammate):} The "Man-Computer Symbiosis" \cite{licklider1960} implies a co-equal, bi-directional partnership. The "colleague"  is an entity you *interact with*, one that has agency, that "supplements your own" competence , and that joins you in "cooperative" decision-making. This path leads directly to the development of conversational agents, "intelligent autonomous teammates" , and the complex, unresolved challenges of shared agency, trust calibration, and accountability.

    \item \textbf{Engelbart's Path (The Tool):} The "Augmenting Human Intellect" framework \cite{engelbart1962} implies a human-centric, human-controlled system. The "H-LAM/T system"  is a system you *inhabit*, an environment you control. The "artifacts" are extensions of the human, not partners. This path leads directly to integrated development environments, graphical user interfaces, and the "supertool" metaphor.
\end{itemize}

This is not merely a semantic difference; it is a fundamental architectural choice. The debate over whether an AI should be an autonomous "teammate" or a powerful "supertool" is the modern echo of the Licklider-Engelbart schism from the 1960s. As will be explored, this tension finds its most vocal modern proponent in Ben Shneiderman, who explicitly rejects the Licklider-inspired "teammate" metaphor in favor of an Engelbart-inspired "supertool". This historical continuity, spanning over 60 years, provides the central narrative for understanding the field's current paradoxes and performance failures.

% \clearpage

% ------------------------------------------------------------------
% --- Section 2: MODERN FRAMEWORKS
% ------------------------------------------------------------------
\section{Modern Frameworks for Human-AI Interaction}

\subsection{The Modern Lexicon: Augmented, Symbiotic, and Collective Intelligence}

The foundational ideas of Licklider and Engelbart have evolved into a modern lexicon, though the field is fraught with terminological confusion. The terms "Augmented Intelligence" and "Symbiotic Intelligence" are often used interchangeably, yet they carry the distinct DNA of their progenitors.

\textbf{Augmented Intelligence (AuI)} is most often framed as the direct successor to Engelbart's vision. It is defined as a "collaborative partnership between humans and AI, designed to enhance our capabilities and support our journey". The focus is squarely on integrating human intelligence (HI) and AI to "harness their strengths and mitigate their weaknesses" \cite{simon2024integrating}. In this model, the human remains the clear principal agent, responsible for "guiding and training" the AI and "driv[ing] AI in the right way". This is an empowerment model.

\textbf{Symbiotic Intelligence}, by contrast, often carries the "bi-directional influence" of Licklider's original concept. While some sources define it almost identically to augmentation , more nuanced definitions describe an interdependence where humans "shape the capabilities, goals, and ethical frameworks of AI agents... while AI agents, in turn, could influence human decision-making, societal norms, and operational practices". This framework implies "mutual adaptation" , a concept closer to Licklider's "living together" than Engelbart's "amplification."

This terminological chaos is a significant finding in itself. The literature is critically inconsistent. For example, some analyses group these terms together , while others issue explicit contradictions, stating that "Hybrid intelligence should not be confused with related concepts such as... symbiotic intelligence". This confusion has practical design consequences. A team that believes it is building an "augmented" system (a tool) will prioritize human control, agency, and explainability. A team that believes it is building a "symbiotic" system (a partner) may prioritize AI autonomy, bi-directional adaptation, and conversational fluency. This report will, therefore, adhere to a clear distinction: *Augmentation* (Engelbart's legacy) as a human-centric tool model, and *Symbiosis* (Licklider's legacy) as a bi-directional, co-adaptive partner model.

A third, emerging framework scales these concepts to the group level: \textbf{AI-Enhanced Collective Intelligence (CI)}. This framework moves beyond the individual human-AI dyad to explore how AI can function within *human collectives*. In this model, the AI's role can vary "from an assistive tool to a participatory member" of the group. The goal is to create a socio-technical system whose "collective intelligence" surpasses that of "either humans or AI in isolation". This is achieved by designing AI systems that "amplify" and "champion human strengths" at the group level, such as "judgment, empathy, [and] creativity".

\subsection{Shneiderman's Human-Centered AI (HCAI) Framework}

The most comprehensive and forceful modern articulation of the Engelbart-style "augmentation" vision is found in Ben Shneiderman's work on "Human-Centered AI" (HCAI) \cite{shneiderman2022}. This framework is presented as a "road map for successful, reliable systems" that bridges "ethical considerations and practical realities".

The core goal of HCAI is to "augment and enhance humans' lives" by creating AI-infused systems that are "reliable, safe, and trustworthy". Shneiderman's philosophy is a "second Copernican Revolution" that "reverses the current emphasis on algorithms and AI methods, by putting humans at the center of systems design thinking". It is a framework built to support "human self-efficacy, promote creativity, clarify responsibility, and facilitate social participation".

\subsubsection{The Two-Dimensional HCAI Framework}

Shneiderman's central theoretical contribution is the \textbf{Two-Dimensional HCAI Framework} \cite{shneiderman2020hcai}. This model is revolutionary because it explicitly *rejects* the perceived "automation-versus-control trade-off". For decades, designers have operated on the assumption that increasing automation necessarily means decreasing human control. Shneiderman reframes this as a false dichotomy.

The HCAI framework consists of two independent axes :
\begin{enumerate}
    \item \textbf{Level of Human Control} (ranging from low to high)
    \item \textbf{Level of Computer Automation} (ranging from low to high)
\end{enumerate}

The design goal is to create systems that reside in the **upper-right quadrant** of this new two-dimensional space: systems that feature "high levels of human control AND high levels of automation".

This simple expansion "is already helping designers imagine fresh possibilities". For example, a word processor with a spell-checker is low automation, low control. A fully autonomous "auto-pilot" for writing would be high automation, low control. An "intelligent teammate" that keeps interrupting is low automation, low control. Shneiderman's upper-right quadrant, however, describes a system like a modern digital camera or a powerful graphic design tool: it uses *immense* levels of internal automation (e.g., for light sensing, image stabilization, layer processing) but provides the human operator with *granular* and *immediate* control over the final creative output. This framework moves the goal from "emulating humans" to "empowering people".

\subsubsection{The "Supertool" Metaphor}

Shneiderman's HCAI framework directly confronts the foundational "Tool vs. Teammate" tension introduced in Section 1. He argues that the Licklider-inspired "teammate" metaphor is not only inaccurate but actively harmful to good design.

Shneiderman explicitly pleads for "a shift in language, imagery, and metaphors away from portrayals of intelligent autonomous teammates towards descriptions of powerful tool-like appliances and tele-operated devices".

This is a direct, modern refutation of Licklider's "colleague". In the HCAI vision, AI is not a partner; it is a "supertool". This philosophical stance has profound implications for design:
\begin{itemize}
    \item \textbf{Teammates} imply conversation, shared goals, and distributed agency. This path leads to ambiguity in responsibility, the risk of "algorithm aversion" or "automation bias," and designs that try to mimic human sociality.
    \item \textbf{Supertools} imply control, reliability, and human agency. This path leads to designs that prioritize "clarify[ing] responsibility" , providing predictable results, and enhancing the human's "self-efficacy".
\end{itemize}

Shneiderman's HCAI, therefore, represents the full maturation of Engelbart's 1962 "augmentation" framework, updated for the age of modern machine learning. The following table (Table \ref{tab:frameworks}) synthesizes this 60-year conceptual arc.

\begin{table}[ht]
\caption{A Comparative Analysis of Foundational Interaction Frameworks}
\label{tab:frameworks}
\centering
\begin{tabular*}{\textwidth}{@{\extracolsep{\fill}} p{0.2\textwidth} p{0.2\textwidth} p{0.25\textwidth} p{0.2\textwidth} p{0.15\textwidth}}
\toprule
\textbf{Framework} & \textbf{Proponent(s)} & \textbf{Primary Goal} & \textbf{Governing Metaphor} & \textbf{Implied AI Role} \\
\midrule
Man-Computer Symbiosis & J.C.R. Licklider (1960) & Facilitate \textit{formulative thinking}; cooperative decision-making. & Biological Symbiont; Organism & Co-equal Partner; Colleague \\
\addlinespace
Augmenting Human Intellect & Douglas Engelbart (1962) & Increase human \textit{capability} to solve complex problems; gain comprehension. & Integrated System (H-LAM/T) & Artifact; System Component \\
\addlinespace
Human-Centered AI (HCAI) & Ben Shneiderman (2022) & Create reliable, safe, and trustworthy tools that enhance human self-efficacy. & Supertool; Appliance & Tool; Tele-operated Device \\
\bottomrule
\end{tabular*}
\end{table}

% \clearpage

% ------------------------------------------------------------------
% --- Section 3: MECHANISMS FOR EFFECTIVE SYMBIOSIS
% ------------------------------------------------------------------
\section{Mechanisms for Effective Human-AI Symbiosis}

\subsection{Introduction: The Prerequisites for Partnership}

Regardless of whether the ultimate design goal is a "teammate" (Licklider) or a "supertool" (Shneiderman), any effective human-AI collaboration requires a degree of mutual understanding. The human must understand what the AI is, what it is doing, and why; the AI must be able to receive and integrate human knowledge, goals, and feedback.

This mutual understanding does not emerge spontaneously. It must be engineered. This Section argues that a successful partnership is built upon three pillars:
\begin{enumerate}
    \item \textbf{Explainable AI (XAI):} The technical mechanism for surfacing the AI's internal logic.
    \item \textbf{Shared Mental Models (SMMs):} The psychological state of mutual understanding that XAI aims to build.
    \item \textbf{Co-Adaptation:} The dynamic learning process through which XAI builds and refines SMMs over time.
\end{enumerate}
Underpinning all three pillars is the foundational concept of \textbf{Trust}. Building trust is "foundational"  to the entire enterprise. It requires the AI to demonstrate reliability, to align with "human-defined goals," and to operate with transparency. Without trust, no true symbiosis or augmentation is possible.

\subsection{Pillar 1: Explainable AI (XAI) as the Mechanism for Trust}

The "black box" nature of modern machine learning models, especially deep learning, is the single greatest barrier to trust. Explainable AI (XAI) is a direct response to this challenge. The U.S. Defense Advanced Research Projects Agency (DARPA) defines XAI as a "suite of machine learning techniques" that, first, "produce more explainable models" while maintaining high performance, and second, "enable human users to understand, appropriately trust, and effectively manage" their AI partners \cite{darpa2016xai}.

XAI is the technical key to "unlock[ing] the black box"  and is frequently cited as a prerequisite for "Human-AI Symbiosis". Its function is to provide the "transparent reporting of human-AI collaborative processes"  and to surface the "why" behind an AI's recommendation. This transparency is what allows humans to "retain ethical authority over interpretive and moral judgments" , rather than blindly accepting algorithmic output.

\subsection{Pillar 2: Shared Mental Models (SMMs) as the Psychological Goal}

XAI is the technical mechanism, but it is not the end goal. The *purpose* of an explanation is to build understanding. In the context of team performance, this understanding is formalized as a "Shared Mental Model" (SMM).

A mental model is an "abstract long-term knowledge structure" that humans use to "describe, explain, and predict the world around them". In a team, an SMM represents the *overlap* of these individual models, creating a shared understanding of the task, the tools, and the team itself. Decades of research in human-human teams has "linked SMM quality to improved team performance" \cite{mathieu2000smm}. An SMM allows team members to "coordinate their behavior" and "account for each other's intentions, beliefs, and behavior" without constant, explicit communication.

This concept is now being extended to human-AI teams, where the AI is considered a "team member with agency" \cite{hoffman2018xai}. For a human-AI team to be effective, they must develop a shared understanding across four key domains, as identified in reviews of the SMM literature :
\begin{enumerate}
    \item \textbf{Task Model:} Shared understanding of "actions and procedures relevant to the task at hand" (e.g., *What is our goal?*).
    \item \textbf{Equipment Model:} Shared understanding of the "equipment and system itself" (e.g., *How does the AI work? What are its limitations?*).
    \item \textbf{Team Model:} Shared knowledge of the "knowledge and skills of team members" (e.g., *What is the human good at? What is the AI good at?*).
    \item \textbf{Interaction Model:} Shared understanding of the "roles and responsibilities" and interaction protocols (e.g., *Who does what? When should I intervene?*).
\end{enumerate}

\subsubsection{The XAI-SMM Causal Chain}

A critical synthesis of the literature reveals that XAI and SMMs are not parallel concepts; they are sequential parts of a single causal chain. \textbf{XAI is the technical input, and an SMM is the psychological output.}

Research in this area explicitly states that the goal of XAI is to "validate that such a process supports the formation of shared mental models". Indeed, the SMM framework provides a new, robust definition of "explainability" itself. Hoffman et al. \cite{hoffman2018xai} argue that a model is only truly "explainable" if it can produce a "good explanation," which they define as an "accurate expression... in language that is understandable by the \textbf{audience} that helps in achieving the \textbf{purpose}".

These criteria (audience, purpose) map directly to the SMM categories. To be effective, an XAI system cannot simply provide a "feature importance" chart. It must be able to deliver explanations that build all four components of the SMM:
\begin{itemize}
    \item It must explain its \textbf{Task Model} (e.g., "My current objective is to identify all anomalies.").
    \item It must explain its \textbf{Equipment Model} (e.g., "I am flagging this because it is a 3-sigma outlier, but I am known to be over-sensitive to noise in this data type.").
    \item It must explain its \textbf{Team Model} (e.g., "This decision requires ethical judgment, which is your domain. I am providing only a statistical summary.").
    \item It must explain its \textbf{Interaction Model} (e.g., "This is a high-priority alert requiring your immediate intervention.").
\end{itemize}
Without this holistic, SMM-oriented approach to explanation, trust cannot be "appropriately" calibrated, and collaboration will, as Section 4 demonstrates, fail.

\subsection{Pillar 3: Co-Adaptation as the Process for Building SMMs}

Shared Mental Models are not built in a single briefing; they are developed, maintained, and repaired over time through a dynamic process of interaction. In human-AI teams, this process is called co-adaptation. This is the modern, technical realization of Licklider's call for a "tighter coupling"  between human and machine.

\subsubsection{Interactive Machine Learning (IML) and Human-in-the-Loop (HITL)}

At its core, co-adaptation is powered by Interactive Machine Learning (IML). IML reframes machine learning not as a static "train then deploy" pipeline, but as a "co-adaptive process"  that involves "letting users build classifiers"  and designing interfaces that facilitate "interactive machine learning".

The most common model of IML is \textbf{Human-in-the-Loop (HITL)}. In a HITL system, human feedback is used to "evaluate the performance"  and "improve models" in a continuous feedback cycle. This human oversight helps "identify and mitigate bias... which encourages fairness in AI outputs".

A more efficient and powerful version of HITL is \textbf{Active Learning}. In this paradigm, the model "selects the data that it wants to be labeled by a human". This concentrates the expert human's effort on the "hardest or most ambiguous examples," leading to faster and more accurate learning.

\subsubsection{Machine Teaching (MT) and Bidirectional Alignment}

The most advanced form of co-adaptation moves from simple feedback (HITL) to "Machine Teaching" (MT) and "Bidirectional Alignment." This is a "bilateral process"  of mutual learning.

First, \textbf{Human Teaches AI (MT)}. The human acts as a "teacher" to the AI "student," iteratively revising their "teaching approach" to align the model with their goals and values.

Second, \textbf{AI Teaches Human (Mutual Adaptation)}. This is the crucial, symbiotic part of the loop. The AI's outputs, explanations, and feedback "in turn, could influence human decision-making, societal norms, and operational practices".

\subsubsection{The Co-Adaptive Loop}

True symbiosis is this dynamic, closed learning loop. The explicit goal of this process is to create "co-adaptive systems that can evolve along with users' mental models". The human's mental model of the AI's capabilities changes as they teach it, and the AI's model of the human's goals changes as it learns.

A perfect, concrete example of this loop is the "After-Action Explanation (AAE) Tool" developed for human-AI teaming testbeds. The AAE workflow is as follows:
\begin{enumerate}
    \item The human and AI agents perform a collaborative task (e.g., a building task in Minecraft).
    \item After the task, they "engage in a group debrief session" using the AAE tool.
    \item The AAE tool, a form of XAI, allows them to "discuss and understand the mission outcomes, agent behaviors, and decision-making processes."
    \item This entire process "facilitate[s] the development of shared mental models" between the human and AI teammates.
\end{enumerate}
This "After-Action Review" loop is the *process* by which XAI (the tool) builds SMMs (the psychological state) over time. This is the modern, practical implementation of the "tighter coupling" and "intimate association" that Licklider envisioned in 1960.

% \clearpage

% ------------------------------------------------------------------
% --- Section 4: THE PERFORMANCE PARADOX
% ------------------------------------------------------------------
\section{The Performance Paradox: Evaluating Human-AI Teams}

\subsection{Introduction: The Paradox of Collaboration}

While Sections 1 through 3 outlined the 60-year philosophical and technical *promise* of human-AI symbiosis, this Section confronts its sobering empirical *performance*. The central challenge to the symbiotic vision is a stark "performance paradox" revealed in recent, large-scale studies.

A systematic review and meta-analysis of human-AI collaboration by Vaccaro, Almaatouq, and Malone \cite{vaccaro2024}, which synthesized the findings from 106 experimental studies and 370 effect sizes , reveals a startling and counter-intuitive bifurcation: in some domains, human-AI collaboration is a clear success, while in others, it is a consistent failure. This Section will deconstruct this paradox, which forms the central empirical problem this report must explain.

\subsection{Negative Synergy: The Failure of Human-AI Decision-Making}

The "most surprising finding"  of the meta-analysis by Vaccaro et al. \cite{vaccaro2024} was that, on average, "human-AI combinations performed significantly worse than the best of humans or AI alone" \cite{vaccaro2024}.

This performance *loss*, or "negative synergy," was not uniform across all tasks. It was heavily concentrated in \textbf{decision-making tasks}. These are tasks that require a human and AI to collaborate on a judgment, such as:
\begin{itemize}
    \item Forecasting demand
    \item Diagnosing medical issues 
    \item Financial decision-making 
    \item Detecting fraudulent content
\end{itemize}

The canonical example from the study is the task of detecting fake hotel reviews. The results were:
\begin{itemize}
    \item \textbf{AI Alone:} 73\% accuracy
    \item \textbf{Human-AI Team:} 69\% accuracy
    \item \textbf{Human Alone:} 55\% accuracy
\end{itemize}
In this case, the addition of a human operator *degraded* the performance of the superior AI. The researchers "hypothesized that because people were less accurate... than the AI, they were also not very good at deciding when to trust the algorithms and when to trust their own judgment". This failure of "trust calibration" resulted in a combined performance lower than that of the AI working in isolation. This finding is not an anomaly; it was the consistent trend across the majority of decision-making tasks studied.

\subsection{Positive Synergy: The Success of Human-AI Content Creation}

In stark contrast, the *same* meta-analysis found "significantly greater gains" and a "positive... effect size" for tasks that involved \textbf{creating content} \cite{vaccaro2024}.

While these tasks represented a smaller fraction of the studies reviewed (only 10\% of papers) , the results were unambiguously positive. In these domains, the human-AI team "were often better than the best of humans or AI working independently".

These "creation" tasks include a wide range of generative and formative activities, such as:
\begin{itemize}
    \item Generating new content and imagery 
    \item Answering questions in a chat 
    \item Software development and coding 
    \item Academic and creative writing 
    \item Design and brainstorming 
\end{itemize}

Case studies in the era of generative AI validate this finding. In coding, AI assistants like GitHub Copilot act as a "brainstorming partner," suggesting real-time code snippets and helping developers "write cleaner code faster". In academic writing, AI tools create a "symbiotic relationship" by enhancing the human's working memory, accelerating drafting, and broadening idea generation. In business, retailers like Walmart and Target have successfully deployed generative AI tools to help employees "more effectively and confidently address questions from customers and colleagues".

This success in creative and generative tasks aligns with the "supertool" metaphor (Section 2), where the AI functions as an augmentation to "amplify human capabilities"  rather than as a partner in a high-stakes judgment. Table \ref{tab:paradox} crystallizes this performance paradox.

\begin{table}[ht]
\caption{Summary of Meta-Analysis Findings on Human-AI Synergy \cite{vaccaro2024}}
\label{tab:paradox}
\centering
\begin{tabular*}{\textwidth}{@{\extracolsep{\fill}} p{0.2\textwidth} p{0.3\textwidth} p{0.2\textwidth} p{0.3\textwidth}}
\toprule
\textbf{Task Domain} & \textbf{Key Finding (Vaccaro et al.)} & \textbf{Synergy Effect} & \textbf{Canonical Examples} \\
\midrule
\textbf{Decision-Making} & "Human-AI combinations performed significantly worse than the best of humans or AI alone." \cite{vaccaro2024} & \textbf{Negative Synergy} (Performance Loss) & Fake review detection; Medical diagnosis; Financial forecasting.  \\
\addlinespace
\textbf{Content Creation} & "Found... significantly greater gains... positive and significantly greater [effect size]." \cite{vaccaro2024} & \textbf{Positive Synergy} (Performance Gain) & Generative writing; Coding; Answering chat questions; Image generation.  \\
\bottomrule
\end{tabular*}
\end{table}

\subsection{Case Study in Scientific Discovery: The "Centaur" Foundation Model}

This performance paradox—failure in judgment, success in creation—is perfectly illustrated in a recent, groundbreaking case study in scientific discovery: the "Centaur" foundation model \cite{binz2024centaur}. This case study not only reinforces the findings of the Vaccaro meta-analysis but, critically, brings the 60-year-old symbiotic vision full circle, demonstrating a modern realization of Licklider's original goal.

"Centaur" is a foundation model of human cognition. It was trained on "Psych-101," a massive dataset of human choices in psychological experiments, to predict and simulate human behavior.

\subsubsection{The Process of Model-Guided Scientific Discovery}

The researchers behind Centaur published a case study of "model-guided scientific discovery" that provides a template for successful, high-level human-AI collaboration.

The process was not to have the AI *make a decision*. Instead, the AI was used as a *collaborator in formulation*:
\begin{enumerate}
    \item \textbf{Start with Human Model:} The researchers had an existing, human-developed cognitive model for a multi-attribute decision-making task.
    \item \textbf{Use AI as Reference:} They used Centaur as a "reference model". They specifically looked for discrepancies, i.e., data points "for which Centaur makes accurate predictions but the [human-discovered] model does not".
    \item \textbf{Refine and Formulate:} This process, which they term "scientific regret minimization" , allowed the human researchers to "guide the development" and "refine" their *own* model.
\end{enumerate}
The end result was a "domain-specific cognitive model that is as predictive as Centaur yet still interpretable". The human did not simply accept the AI's "decision"; they used the AI's "competence" to *augment their own understanding* and *formulate a new, better theory*.

\subsubsection{The Realization of Licklider's Vision}

This Centaur case study is arguably the perfect modern realization of J.C.R. Licklider's 1960 vision. Licklider's primary, foundational goal was "to let computers facilitate \textbf{formulative thinking}". This is *exactly* what the Centaur model achieved. The researchers were, in Licklider's own words, "thinking in interaction with a computer in the same way that you think with a colleague whose competence supplements your own".

The success of this case study, in a creative/discovery task, reinforces the "Positive Synergy" finding from Vaccaro et al. \cite{vaccaro2024}. It suggests that the true future of human-AI symbiosis lies not in automating *judgment* (which fails), but in augmenting *formulation* (which succeeds). The paradox is resolved: the "teammate" model fails in decision tasks, but the "supertool" (or "colleague") model triumphs in creative and scientific ones. The next Section will explore *why* the decision-making model fails so catastrophically.

% \clearpage

% ------------------------------------------------------------------
% --- Section 5: PSYCHOLOGICAL AND ETHICAL CHALLENGES
% ------------------------------------------------------------------
\section{Psychological and Ethical Challenges in Symbiosis}

\subsection{Introduction: Why Do Human-AI Decision Teams Fail?}

The performance paradox detailed in Section 4 demands a deeper explanation. Why do human-AI teams consistently *fail* in decision-making tasks, exhibiting negative synergy \cite{vaccaro2024}? The hypothesis from the Vaccaro meta-analysis was a failure of trust calibration. This Section will dissect that failure, arguing that the failures are not merely technical but deeply psychological. They stem from a fundamental mismatch between the design of "Algorithm-in-the-Loop" systems and the realities of human cognition.

\subsection{The "Algorithm-in-the-Loop" (AIL) Problem}

The seminal work in this area is "The Principles and Limits of Algorithm-in-the-Loop Decision Making" by Ben Green and Yiling Chen \cite{green2019principles}. The AIL process, where "machine learning models inform people" , is the dominant paradigm for human-AI decision-making, particularly in high-stakes fields like medicine, finance , and criminal justice.

Green and Chen argue that this entire sociotechnical system is flawed because "people struggle to interpret machine learning models and to incorporate them into their decisions". Their experimental study  evaluated whether human-AI teams could satisfy three basic principles of ethical decision-making: accuracy, reliability, and fairness. While they found that AI assistance could improve *accuracy* (Desideratum 1), they found a total failure on the other two principles.

Under all experimental conditions, participants:
\begin{enumerate}
    \item \textbf{Failed to Evaluate:} They "were unable to effectively evaluate the accuracy of their own or the risk assessment's predictions".
    \item \textbf{Failed to Calibrate:} They "did not calibrate their reliance on the risk assessment based on the risk assessment's performance".
    \item \textbf{Exhibited Bias:} They "exhibited bias in their interactions with the risk assessment".
\end{enumerate}

\subsubsection{The Causal Link to Performance Failure}

The AIL problem, as defined by Green and Chen \cite{green2019principles}, provides the causal mechanism that *explains* the "Negative Synergy" finding from the Vaccaro meta-analysis \cite{vaccaro2024}.
\begin{itemize}
    \item The Vaccaro study *hypothesized* that human-AI decision teams fail because humans "are not very good at deciding when to trust".
    \item The Green \& Chen study *proved* it, demonstrating empirically that humans *cannot* "effectively evaluate" or "calibrate their reliance".
\end{itemize}

This is the core of the failure. The human in the loop, who is theoretically present to provide oversight and correct AI errors, is cognitively incapable of doing so reliably. Unable to properly "evaluate the accuracy" of the AI's advice, the human injects *noise* rather than *signal* into the decision. They over-trust a wrong AI, or under-trust a correct one. In either case, the human's presence drags the team's combined performance below that of the AI working alone. This is a direct, measurable consequence of a failure to build the Shared Mental Models (SMMs) described in Section 3.

\subsection{The Spectrum of Cognitive Bias}

The AIL problem manifests as a set of specific cognitive failure modes. The human operator, lacking the SMMs to "appropriately trust"  the AI, defaults to one of two cognitive errors. They are trapped on an unstable spectrum between aversion and bias.

\subsubsection{Algorithm Aversion}

On one end of the spectrum is \textbf{Algorithm Aversion}. This is a "biased negative assessment of algorithmic systems"  or a simple "algorithm aversion". Even when presented with evidence of an algorithm's superior performance, humans may distrust it, preferring their own (often inferior) intuition. This leads to the human ignoring a correct AI prediction, thus degrading team performance.

\subsubsection{Automation Bias and Complacency}

On the opposite end of the spectrum is \textbf{Automation Bias}. This is an "overreliance on automated systems"  where a human "uncritically" defers to a technology they "deem more proficient than themselves". This is often the more "dangerous" of the two errors.

This "automation complacency"  is particularly pernicious when the AI model itself is flawed or biased. For example, "if a hiring algorithm is biased against minorities... the human-in-the-loop who is responsible for oversight" may, due to automation bias, "perpetuate or even exacerbate the already existing bias instead of mitigating it". In this scenario, the human is not a failsafe; they are an amplifier for the AI's error.

\subsubsection{The Unstable Spectrum}

The human operator is trapped on this unstable spectrum. Lacking the deep, SMM-based understanding of the AI's "Equipment Model"  or "Team Model" , they are forced to use simple, flawed heuristics. They default to either "algorithm aversion" (distrust) or "automation bias" (blind trust), both of which lead to performance degradation and systemic failure.

\subsection{The Long-Term Risk: Cognitive Deskilling}

Beyond the immediate performance failures of a single decision, the long-term, persistent use of AIL systems presents a more insidious risk: \textbf{cognitive deskilling}.

When human operators "offload" their cognitive work to an AI, it can lead to the "erosion of diagnostic reasoning and clinical judgment"  or the dilution of their own creative "voice". This is not just a risk of "deskilling" ; it is a risk of fundamentally altering human cognition.

A 2021 study by Fügener et al., cited in , described this as the \textbf{"Borgs" Phenomenon}. The study's model found that when working with AI, humans can "lose their unique knowledge, defined as knowledge that the AI does not have, thus, become 'borgs'". The AI transforms human cognition, leaving the human "unaware about this". This leads to major concerns about the "erosion of tacit knowledge transfer"  and "value misalignment"  as human expertise atrophies through disuse.

\subsection{Ethical Governance: Accountability, and Transparency}

This web of psychological failures creates an ethical and governance nightmare. The entire "human-in-the-loop" model is often proposed as a solution to AI accountability, but the empirical evidence shows it is a flawed premise.

If the human cannot "calibrate their reliance"  and is prone to "exacerbate" AI bias , then who is truly responsible for a flawed outcome? The "bi-directional influence"  of a symbiotic system blurs the "clear lines of responsibility". This creates an "accountability gap".

The clear consensus is that the solution must be twofold:
\begin{enumerate}
    \item \textbf{Technical Transparency:} We must "unlock the black box"  through robust XAI systems (as described in Section 3).
    \item \textbf{Ethical Governance:} We must "champion accountability"  by designing systems that, like Shneiderman's HCAI, "clarify responsibility"  from the outset, preserving "human accountability for ethical and interpretive judgments".
\end{enumerate}
Without such a framework, the "iterative and contingent" process of building trust  is impossible, and the symbiotic vision is doomed to fail.

% \clearpage

% ------------------------------------------------------------------
% --- Section 6: CONCLUSION
% ------------------------------------------------------------------
\section{Conclusion: Toward a Future Framework for Human-XAI Symbiosis}

\subsection{Summary of Findings: The Bifurcation of the Symbiotic Vision}

This report has charted the 60-year quest for human-machine partnership, a journey that began with the foundational, and fundamentally competing, visions of J.C.R. Licklider \cite{licklider1960} and Douglas Engelbart \cite{engelbart1962}. This journey, which created the "teammate vs. tool" tension that defines the field, has led to a fundamental, data-driven paradox in the modern era.

The analysis concludes that the symbiotic vision has bifurcated:
\begin{enumerate}
    \item \textbf{The Failure of the "Teammate" in Judgment.} The Licklider-inspired "teammate" model, in which humans and AI partner to make \textbf{decisions}, is empirically failing. The large-scale meta-analysis by Vaccaro et al. \cite{vaccaro2024} shows this combination produces *negative synergy*, performing worse than AI alone. This report has identified the causal mechanism for this failure: the "Algorithm-in-the-Loop" (AIL) problem \cite{green2019principles}. As shown by Green \& Chen, humans are psychologically ill-equipped to "evaluate... accuracy" or "calibrate... reliance" on an external AI partner. Trapped on the unstable spectrum between "algorithm aversion"  and "automation bias" , the human becomes a source of cognitive noise, all stemming from a failure to build robust Shared Mental Models (SMMs).

    \item \textbf{The Success of the "Supertool" in Formulation.} The Engelbart-inspired "augmentation" model, modernized by Shneiderman as the "supertool" , is empirically succeeding. In tasks of \textbf{content creation} and \textbf{problem formulation}, the Vaccaro et al. \cite{vaccaro2024} meta-analysis finds clear *positive synergy*, where the human-AI team outperforms either member alone. This success was perfectly illustrated by the "Centaur" foundation model case study. In that case, the AI was used not to *make a judgment*, but as a "colleague"  to help researchers *formulate* a new scientific theory. This, critically, is the modern realization of Licklider's *original* and most important goal: "to let computers facilitate \textbf{formulative thinking}".
\end{enumerate}

The resolution of the paradox is, therefore, that Licklider's *primary goal* (formulation) is being achieved through Engelbart's *method* (augmentation).

\subsection{A New Theoretical Lens: Dual-Process Theory and the "Extended Self"}

This report concludes by proposing a path forward, one that synthesizes these findings into a new, more robust theoretical framework for human-XAI symbiosis. A new framework proposed by Litvinova et al. \cite{litvinova2024} suggests that true, long-run symbiosis is grounded in two concepts: the \textbf{notion of the "extended self"} and the \textbf{dual-process theory of cognition}.

Dual-process theory posits that human cognition operates on two tracks: \textbf{Type 1} (autonomous, fast, intuitive) and \textbf{Type 2} (deliberative, slow, analytical). The failures of the AIL problem (Section 5) can be seen as a breakdown in this system: the AI's "suggestion" creates a "cognitive conflict" that the human's Type 2 process is forced to resolve, a process for which it is (as Green \& Chen proved) poorly equipped.

The "extended self" framework \cite{litvinova2024} offers a theoretical solution. In this model, through persistent, regular, and co-adaptive collaboration, the XAI is no longer perceived as an *external agent* to be "trusted" or "distrusted." It becomes integrated into the human's own cognitive framework—an "extended self".

\subsubsection{A Path Beyond the Performance Paradox}

This "extended self" framework offers a profound, theoretical solution to the AIL performance paradox. The failures of aversion and bias  are triggered precisely *because* the human perceives the AI as an *external* teammate, forcing an explicit (and, as we have seen, flawed) "trust calibration" step.

The "extended self" model \cite{litvinova2024} bypasses this cognitive bottleneck. By forming a "unitary human-XAI symbiotic agency" , the XAI's response is no longer an *external suggestion* that must be deliberated by a skeptical Type 2 process. Instead, it is integrated as one of the human's *own* "autonomously generated responses" (a Type 1 intuition) from which deliberation can begin.

This reframes the entire problem. It is no longer "Human vs. AI." It becomes a single, "extended" cognitive system. This is the only way to achieve the deep "cooperative interaction" and "very close coupling"  that J.C.R. Licklider envisioned in 1960.

\subsection{Future Research Directions}

This new framework suggests a clear agenda for future research, moving beyond short-term, lab-based experiments that "force" a trust calibration, and toward studies of long-term co-adaptation. Key directions include:

\begin{itemize}
    \item \textbf{Empirical Validation:} There is an urgent need for "empirical testing of... principles"  derived from the "extended self" framework. This requires longitudinal studies that track how (and if) users' mental models evolve to integrate AI as part of their cognitive "self."
    \item \textbf{Mitigating Deskilling:} Research must focus on the "erosion of tacit knowledge transfer"  and "cognitive deskilling". The most promising path is the development of "hybrid models that pair AI with mentorship" , using AI not as a replacement for, but as a scaffold for, human learning.
    \item \textbf{Designing for Formulation, Not Just Decision:} The findings of Vaccaro et al. \cite{vaccaro2024} and the Centaur case study \cite{binz2024centaur} are a clear mandate. Design efforts should be re-focused from building AI "decision-makers" (which fail) to building "formulation partners" and "supertools" (which succeed).
    \item \textbf{New Psychological Frameworks:} Finally, new psychological and technical frameworks are needed "in which to evaluate the 'fit' of humans or AI in ever-shifting contexts".
\end{itemize}

The future of research "lies not in human versus AI competition, but in thoughtfully designed partnerships that amplify the best of both human and artificial intelligence". By learning from the empirical failures of the AIL model and embracing the success of the "augmentation" model, the field can finally move beyond the "tool vs. teammate" debate and begin to build the "unitary" cognitive systems that represent true symbiosis.

\bibliographystyle{plainurl}
\bibliography{references}
\end{document}